\documentclass[useAMS,usenatbib,usegraphicx,usetimes]{mn2e}
\usepackage{amssymb}
\usepackage{times,color}

\def \<{\langle}
\def \>{\rangle}
\def\acknowledgments {\section*{Acknowledgements}%
  \addtocontents{toc}{\protect\vspace{6pt}}%
  \addcontentsline{toc}{section}{Acknowledgements}%
}

\title[WMAP Timing Error]
{Diagnosing Timing Error in WMAP Data}

\author[H. Liu \& T. P. Li]{Hao Liu$^{1}$\thanks{E-mail: liuhao@ihep.ac.cn}, 
Shao-Lin Xiong$^{1}$ and Ti-Pei Li$^{1,2,3}$\thanks{E-mail: litp@tsinghua.edu.cn}\\
$^{1}$Key Laboratory of Particle Astrophysics, Institute of High Energy Physics,
Chinese Academy of Sciences, Beijing, China\\
$^2$Department of Physics and Center for Astrophysics, Tsinghua University, Beijing, China\\
$^3$Department of Engineering Physics and Center for Astrophysics, Tsinghua University, Beijing, China
}

\begin{document}

\date{}

\maketitle

\label{firstpage}

\begin{abstract}
The Doppler dipole signal dominates the cosmic microwave background (CMB) anisotropy maps obtained by the Wilkinson Microwave Anisotropy Probe (WMAP) mission, and plays a key role throughout the data processing. Previously, we discovered a timing asynchronism of -25.6\,ms between the timestamps of the spacecraft attitude and radiometer output in the original raw WMAP time-ordered data (TOD), which, if not corrected in following data processing, would generate an artificial quadrupole component ($l=2$) in recovered CMB maps (Liu, Xiong \& Li 2010).  Recently,  Roukema (2010b) proves that there does exist a timing-offset-induced error corresponding to about -25.6\,ms in the WMAP calibrated TOD by studying the fluctuation variance per pixel in the temperature map recovered from the TOD as a function of assumed timing-offset.  Here,  we find evidence directly in the WMAP TOD for such an uncorrected timing error, possibly occurred in calculating the Doppler dipole signal during the WMAP team's TOD data processing. The amplitude is highly significant and is consistent with previous work. We also show that the uncorrected timing-offset can lead the WMAP CMB quadrupole to be substantially overestimated.
\end{abstract}


\section{Introduction}
The WMAP mission makes measurements for the CMB with two antennas, A and B, and records in time-order the raw uncalibrated TOD \citep{ben03a,hin03}
\begin{equation}\label{dr}
d_{raw}=g\cdot d+b\,.
\end{equation}
With the instrument gain $g$ and baseline $b$, the calibrated TOD can be obtained as
\begin{equation}\label{d}
d=T_A-T_B+D\,,
\end{equation}
where $T_A$ and $T_B$ are the antenna temperatures of the antennas A and B, respectively, that would be measured if the spacecraft CMB-frame velocity dipole were zero, $D$ is the Doppler dipole signal induced by the motion of spacecraft
\begin{equation}\label{dip}
D = \frac{T_0}{c}\, {\bf v} \cdot ({\bf n}_{_A} - {\bf n}_{_B})
\end{equation}	 				
with $T_0=2.725$ K being the CMB monopole, $c$ the speed of light, $\bf{v}$ the velocity of the observer relative to the CMB rest frame, ${\bf n}_{_A}$ and ${\bf n}_{_B}$ the unit direction vectors of the antenna A and B respectively.\footnote{The exact form of Eq.~\ref{dip} should consider the transmission imbalance, as given by Eq. 3 of ~\citet{hin09}. We have adopted the exact form in our data analysis; however, it's also confirmed that this simplified form is accurate enough in this work. The Sun velocity to the CMB rest frame we used is $(-26.26, -243.71, 274.63)$ km/s in the Galactic coordinate.} The amplitude of the Doppler dipole signal is about 3\,mK, nearly two order of magnitude stronger than the CMB anisotropy ($\sim 50\,\mu$K). For sky map-making, the dipole signal has to be subtracted from the calibrated TOD to get the dipole-subtracted TOD
\begin{equation}\label{ds}
d_s=d-D\,.
\end{equation}

The calibration parameters $g$ and $b$ in Eq.\,\ref{dr} are determined by a dipole-based calibration procedure \citep{hin03},  where the Doppler dipole signal is initially taken as a standard\footnote{The calibration is then further improved by iteratively solve for the dipole and the map.} which can be calculated by Eq.\,\ref{dip} from the spacecraft direction and velocity data. Thus an error in evaluating the dipole signal could be expected to induce the calibration parameters in error, and then twist the calibrated TOD. Furthermore, a small error in evaluating the Doppler dipole may arouse a significant consequence on the final CMB temperature map via the dipole subtraction (Eq. \ref{ds}). For example, in dipole calculation with Eq.\,\ref{dip}, an antenna direction deviation as small as $\sim 7'$,  just about a half-pixel in the usual WMAP resolution with the resolution parameter $N_{side}=512$ \citep{ben03b}, or $\sim 20$\,ms timing asynchronism,  can cause the differential dipole signal and then the dipole-subtracted TOD to be biased by $\sim10-20\,\mu$K \citep{lxl10}. Therefore, the Doppler dipole signal plays a key role throughout the WMAP data processing, and its error should be inspected very carefully.

The WMAP spacecraft is continuously scanning the sky. For each datum $d_{raw}(t)$ or $d(t)$ observed at a provided time $t$ in the Science Data Table of the WMAP TOD archive \citep{lim08}, one has to derive ${\bf n}_{_A}(t)$, ${\bf n}_{_B}(t)$  and ${\bf v}(t)$   from the Meta Data Table and calculate the corresponding Doppler signal by using Eq.\,\ref{dip}. It is obvious that, in both data-calibration and map-making, the time $t$ used to calculate the Doppler signal  $D(t)$ must be synchronous to the time $t$ of the science datum $d_{raw}(t)$ or $d(t)$. But we found  in the WMAP TOD archive that the Meta Data Table is not recorded simultaneously with the Science Data Table: a datum in the Meta Data Table is recorded 25.6\,ms later than the corresponding science datum in the Science Data Table for all bands, in other words, there exists a -25.6\,ms timing-offset between  the Meta Data Table and the Science Data Table (Liu, Xiong \& Li 2010).  Such a timing-offset, if were incorporated into the calibration procedure without being corrected, would induce a significant error in the calibrated TOD and then on the CMB map due to improperly evaluating the Doppler-induced signal $D(t)$. To check the suspected timing-offset-induced error in the official WMAP calibrated TOD,  Roukema (2010b) recently studies the median per map of the temperature fluctuation variance per pixel as a function of assumed timing-offset and proves that the results are consistent with a -25.6\,ms offset at $1.4\sigma$  whereas the hypothesis of no timing error in the WMAP calibrated TOD is rejected at a significance of  $8.5\sigma$. In this work, we use a more direct technique to show the existence of timing-offset-induced error in the official WMAP calibrated TOD.

\section{Checking the Timing Offset}
\subsection{Method}

Let $t_i$ denotes the observation time for the $i$-th datum in the WMAP time-ordered data, $d^{(w)}(t_i)$ the released WMAP calibrated differential datum at $t_i$, $D^{(w)}(t_i)$ the real dipole  component existed in $d^{(w)}(t_i)$. For an assumed timing-offset $\Delta t$, we can compute the dipole component $D(t_i+\Delta t)$ at time $t_i+\Delta t$ by Eq.\,\ref{dip} with ${\bf v}$, ${\bf n}_{_A}$ and  ${\bf n}_{_B}$ all at $t_i+\Delta t$ and obtain the dipole-subtracted TOD
\begin{eqnarray}\label{dsw}
\lefteqn{d_s^{(w)}(t_i;\Delta t)=d^{(w)}(t_i)-D(t_i+\Delta t)} \nonumber \\
& & =[T_A(t_i)-T_B(t_i)]+[D^{(w)}(t_i)-D(t_i+\Delta t)]\,,
\end{eqnarray}
and the statistic
\begin{eqnarray}\label{v}
\lefteqn{V(\Delta t)=\sum_i [d_s^{(w)}(t_i;\Delta t)]^2} \nonumber \\
& & =\sum_i\{ [T_A(t_i)-T_B(t_i)]^2+[D^{(w)}(t_i)-D(t_i+\Delta t)]^2+ \nonumber \\
& & \hspace{8mm} +\, 2 [T_A(t_i)-T_B(t_i)][D^{(w)}(t_i)-D(t_i+\Delta t)]\}\,.
\end{eqnarray}
Since the Doppler dipole signal is causally determined only by the motion of the spacecraft, unrelated to the astrophysical temperature fluctuations, the third term of the sum in the second line of Eq.\,\ref{v}, i.e. the cross-term between $T_A(t_i)-T_B(t_i)$ and $D^{(w)}(t_i) - D(t_i+\Delta t)$ in Eq.\,\ref{v}, summed over $i$, is likely to vary randomly around zero as $\Delta t$ is varied (we postpone further discussion of this term to \S\ref{sec:cons} below).  Without timing error in the WMAP data calibration, the dipole $D^{(w)}(t_i)$ really existed in the calibrated TOD should be identical to that calculated with $\Delta t=0$, thus we have $D^{(w)}(t_i)-D(t_i)=0$ in the right side of Eq.\,\ref{dsw} and $V(\Delta t)$ is on average minimized at $\Delta t=0$.  However, if there exists an uncorrected non-zero timing-offset $\Delta t^*$ between the spacecraft attitude data and science data during the WMAP data processing, then we have $D^{(w)}(t_i)=D(t_i+\Delta t^*)$, and, consequently, $V(\Delta t)$ is on average minimized at $\Delta t=\Delta t^*$. Thus, we can produce Doppler signal sets with different timing-offset $\Delta t$  and compute the statistic $V(\Delta t)$ respectively. The average of all timing-offset $\Delta t^*$ that minimize $V(\Delta t)$ is a proper estimation of the amplitude of suspected timing error in the WMAP data processing.

\subsection{Result}
In the WMAP TOD archive, each science frame contains 15-30 observations, and each observation last for a duration $\tau$ which is 102.4\,ms, 76.8\,ms, and 51.2\,ms for Q-, V- and W-band, respectively. Following the convention used in previous works (Liu, Xiong \& Li 2010; Roukema 2010a, 2010b), the timing-offset used in this work is a relative one in percentage of the duration $\tau$ of each observation
\begin{equation}\label{dtr}
\Delta t_r=(t_D-t_0)/\tau,
\end{equation}
where $t_D$ is the time for the instantaneous Doppler dipole signal $D^{(w)}$, and $t_0$ the starting time of each observation. If the time used for the science data and  spacecraft attitude data are synchronous, then the time for the instantaneous Doppler dipole signal $D^{(w)}$ should be at the center of each observation and $\Delta t_r=0.5$ (same for all bands, neglecting the $\tau$ difference. That's why the relative time $\Delta t_r$ is preferred). Similarly, if the time for the instantaneous Doppler dipole signal $D^{(w)}$ is at the start of each observation (e.g., due to timing offset), then $\Delta t_r=0$, and in case of the end of each observation we have $\Delta t_r=1$. The relationship between $\Delta t$ and $\Delta t_r$ is

\begin{equation}\label{dtdtr}
\Delta t=(\Delta t_r-0.5)\tau\,.
\end{equation}

\begin{figure}
\includegraphics[width=0.25\textwidth]{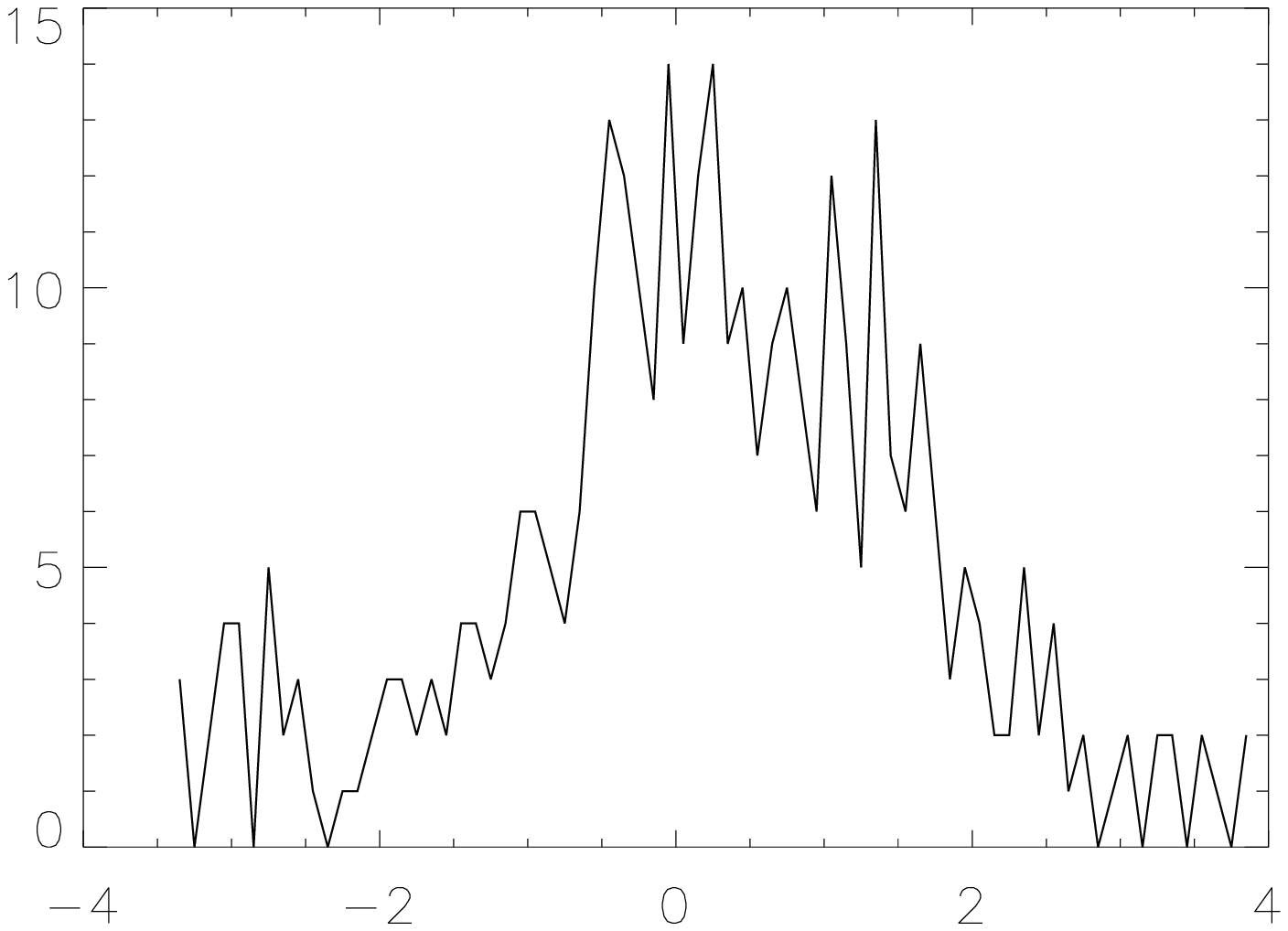}
\includegraphics[width=0.25\textwidth]{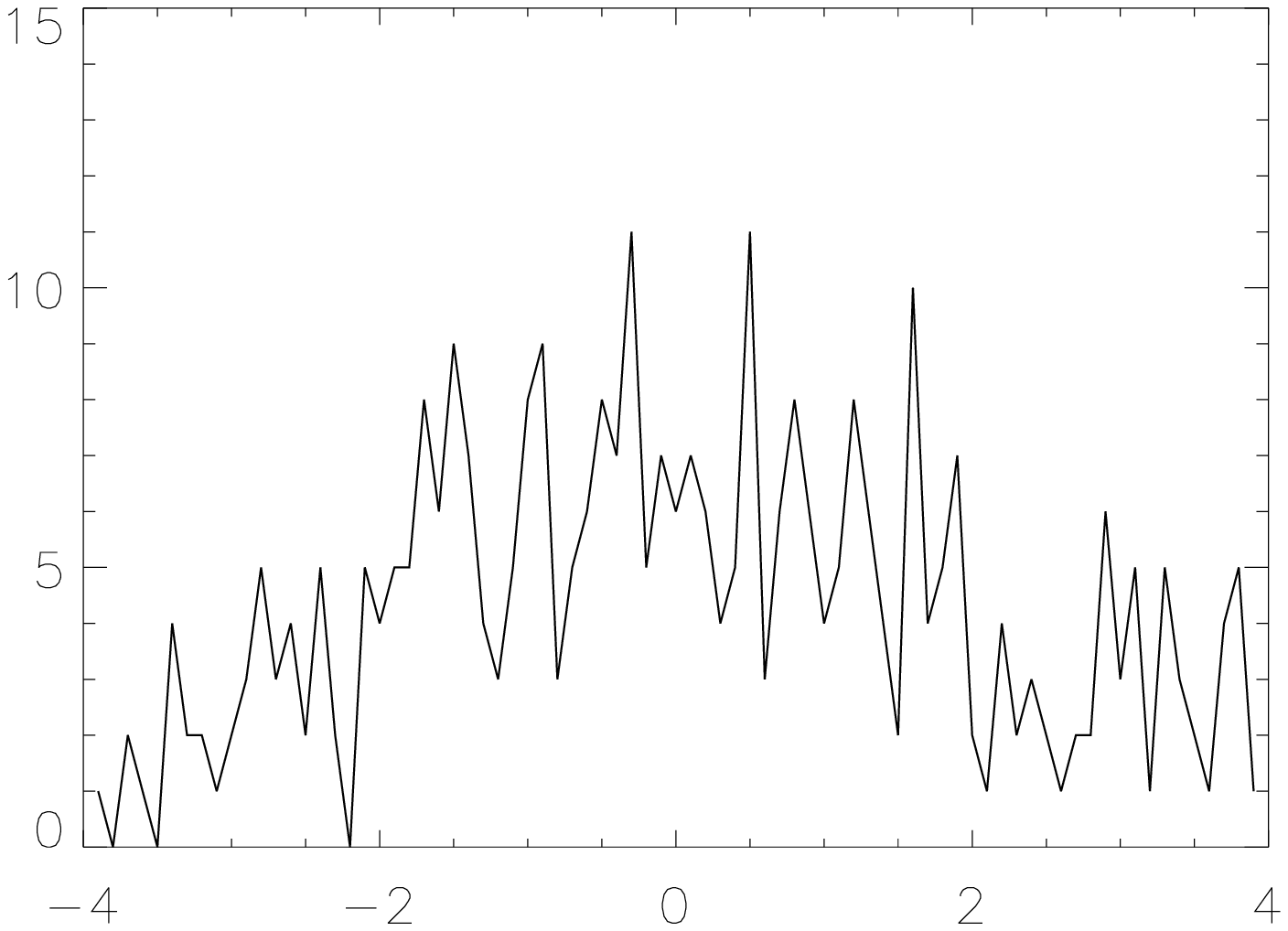}
\includegraphics[width=0.25\textwidth]{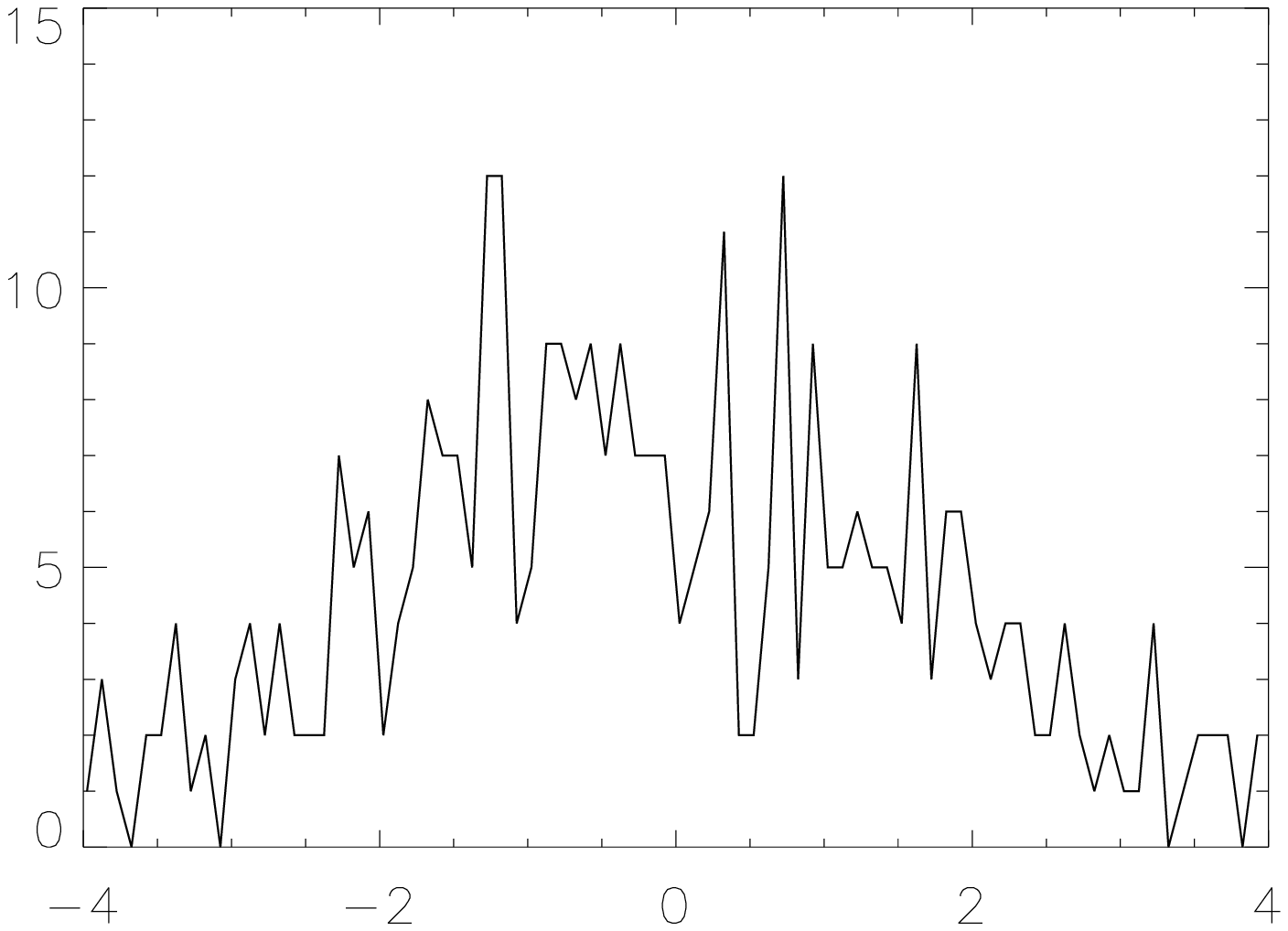}
\caption{The histogram plots of one-year's  averaged $\Delta t^*_r$ before smoothing for the WMAP Q-band (shown in the upper panel), V-band (middle panel), and W-band (bottom panel), respectively.}
\label{fig:hist}
\end{figure}

For reducing the effect of foreground emission, we use the KQ75 mask \citep{gold09} to remove all observations with either side in the mask. In calculating $V(\Delta t_r)$ with Eq.\,\ref{dsw}, Eq.\,\ref{v} and Eq.\,\ref{dtdtr}, the argument $\Delta t_r$ is taken from -6 to +6 with a step of 0.1. For each waveband, we compute for one-day observation periods and record the $\Delta t^*_r$ that minimizes $V(\Delta t_r)$ for each one-day period. For all 7-year TOD, there are about $7\times 365\,\Delta t^*_r$.  Finally, we pick out all  $-4<\Delta t^*_r<4$ (this excludes $3-7\%$ data), compute the average $\<\Delta t^*_r\>$ and the standard error of $\<\Delta t^*_r\>$ by $\sum{(\Delta t^*_r-\<\Delta t^*_r\>)^2}/\sqrt{N(N-1)}$ to give the final relative timing-offset estimation. The obtained $\<\Delta t^*_r\>$ are $0.282\pm0.025,  0.083\pm0.034$, and  $0.085\pm0.032$ for Q-, V- and W-band, respectively.  The corresponding results for the average timing-offset $\<\Delta t^*\>$ are listed in Table\,1, and the histogram plots of one-year's $\Delta t^*_r$ is also shown in Fig.~\ref{fig:hist}.

\begin{table} 
\begin{center}
\begin{tabular}{cc}
 \hline
\hline
Wave Band & $\<\Delta t^*\>$ (ms) \\
\hline
Q   & -22.33$\pm$2.52 \\
V   & -32.06$\pm$2.62 \\
W   & -21.24$\pm$2.30 \\ \hline
All & -24.22$\pm$1.47 \\
\hline
\end{tabular}
\caption{Timing offset in WMAP data. If there is no problem in the official WMAP calibrated TOD, then we should expect $\<\Delta t^*\>=0$\,ms for all bands. However, this is apparently not true, indicating that there is a timing-asynchronism problem in the WMAP data processing.}
\end{center}
\end{table}

In order to test the $\<\Delta t^*_r\>$ estimation, we use a secondary method in which we pick out all $-4<\Delta t^*_r<4$ likewise, then smooth them with a 20-point-window boxcar filter, then count its histogram with 120 bins, and then fit the histogram with Gaussian distribution and take the center of the fitted Gaussian peak as the best-fit $\Delta t^*_r$ estimation. The obtained $\<\Delta t^*_r\>$ for Q-, V- and W-band from the secondary method are 0.282, 0.108 and 0.060, respectively, well consistent to the primary method.

If there is no timing error in the official WMAP calibrated TOD, we should get a nearly zero $\<\Delta t^*\>$ for each band, but from Table\,1 we can see that such a hypothesis is rejected at least at $8.8\sigma$ (estimated by the band with the lowest significance for this) and the existence of a -25.6\,ms asynchronism  in the WMAP calibrated TOD is within $0.94\sigma$ (estimated by the average value in the last line of Table 1), both are consistent with~\citet{rou10b}.

\section{Consequence of the Timing Offset}
\label{sec:cons}
The above result of checking the timing error in the WMAP TOD demonstrates that the official WMAP calibrated TOD $d$ have been incorrectly calibrated because of incorrect usage of the Doppler dipole signal $D$. Ideally, for the uncalibrated science data $d_{\mathrm{raw}}$ we should use $t + \left<\Delta t^*\right>$ instead of $t$ to estimate the Doppler dipole $D$ in Eq.~3 and find $g$ and $b$ in Eq.~1, in order to recalculate the calibrated science data
$d_{\mathrm{recalib}}$, and then derive sky maps from $d_{\mathrm{recalib}}$ without any further timing offset. However, an easier, approximate method on large scales is to use $t + \left<\Delta t^*\right>$ instead of $t$ in making maps from the (incorrectly) calibrated science data $d$ that are publicly available on the WMAP website.

To show the consequence of the timing error upon the official WMAP calibrated TOD, as an example, the quadrupole (power at $l=2$) derived with pseudo-$C_l$ method from the WMAP5 official CMB maps is $112.7\,\mu$K$^2$ (by convention, the given value is $l(l+1)C_l/(2\pi$)), but in the modified map recovered from the new dipole-subtracted TOD $d_s$ with -25.6\,ms timing-offset correction in evaluating the Doppler dipole $D$, the quadrupole decreases down to  $28.6\,\mu$K$^2$ -- the timing error in calibration can generate an artificial quadrupole signal, leading the cosmological CMB quadrupole to be significantly overestimated in the WMAP release\footnote{By using zero timing-offset $\Delta T$, we have obtained fully consistent quadrupole result to the WMAP release (using WMAP5 TOD, KQ85 mask and all Q, V, W bands, and such conditions are same for a none-zero $\Delta T$), indicating that our map-making and quadrupole estimating processes are fine.}.

In previous works (Liu, Xiong \& Li 2010; Moss, Scott \& Sigurdson 2010; Roukema 2010a), it has been discovered and confirmed that, given $\Delta t\sim -25.6$ ms and without using any CMB data, an artificial quadrupole component that is very similar to the released WMAP CMB quadrupole can be produced (Fig.~\ref{fig:compare}).\footnote{In~\citet{moss10}, they have obtained an artificial quadrupole structure that closely resembles ours, but they claim that the amplitude is lower than ours.}   This strongly questions the validity of the WMAP result, and indicates that most of the released WMAP CMB quadrupole might be artificial because of the timing asynchronism effect, as demonstrated by~\citet{rou10b,lxl10,liu11a} and this work.

\section{Discussion}
\label{sec:discussion}
\begin{figure}
\includegraphics[angle=90,width=0.22\textwidth]{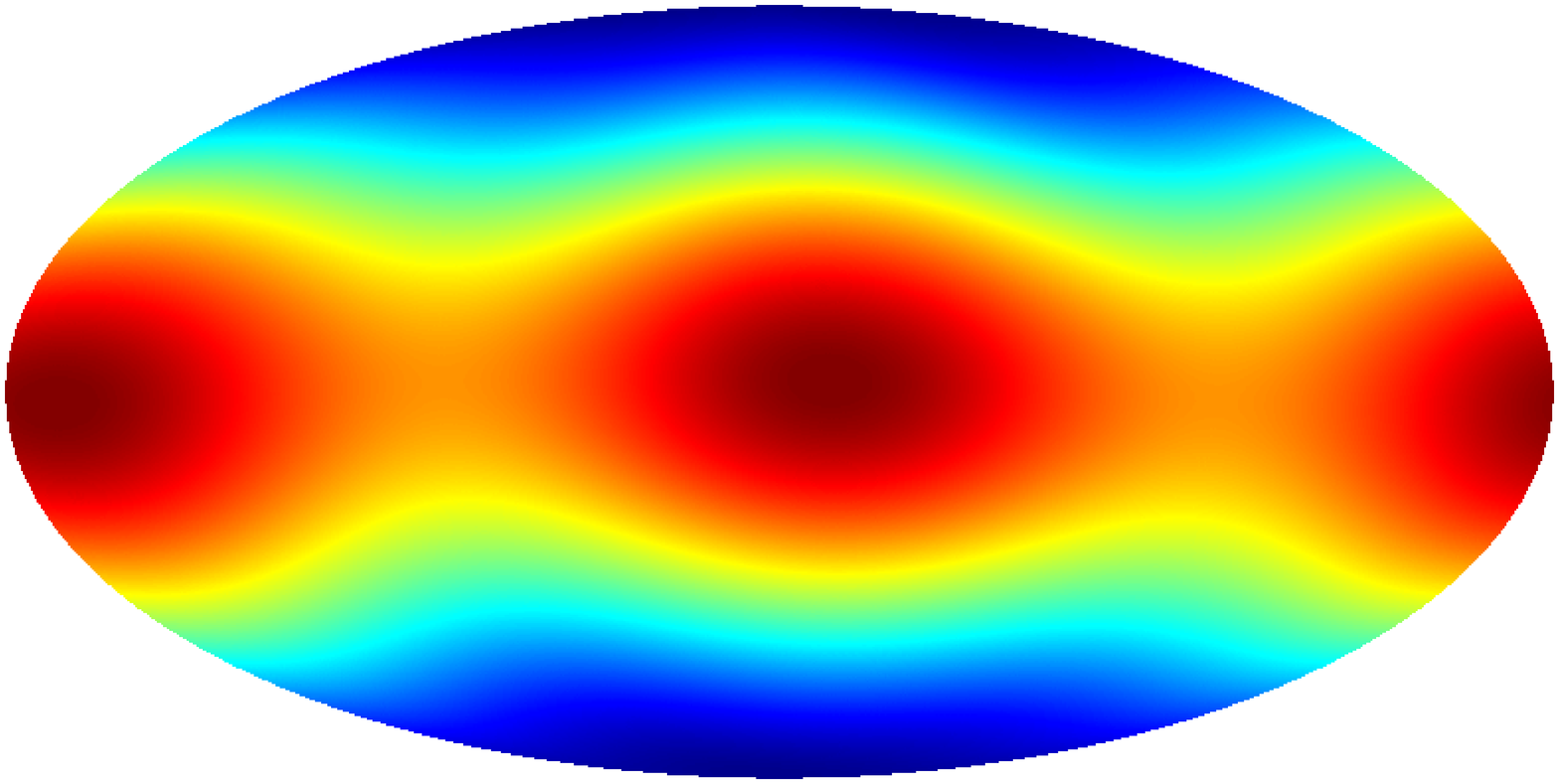}
\includegraphics[angle=90,width=0.22\textwidth]{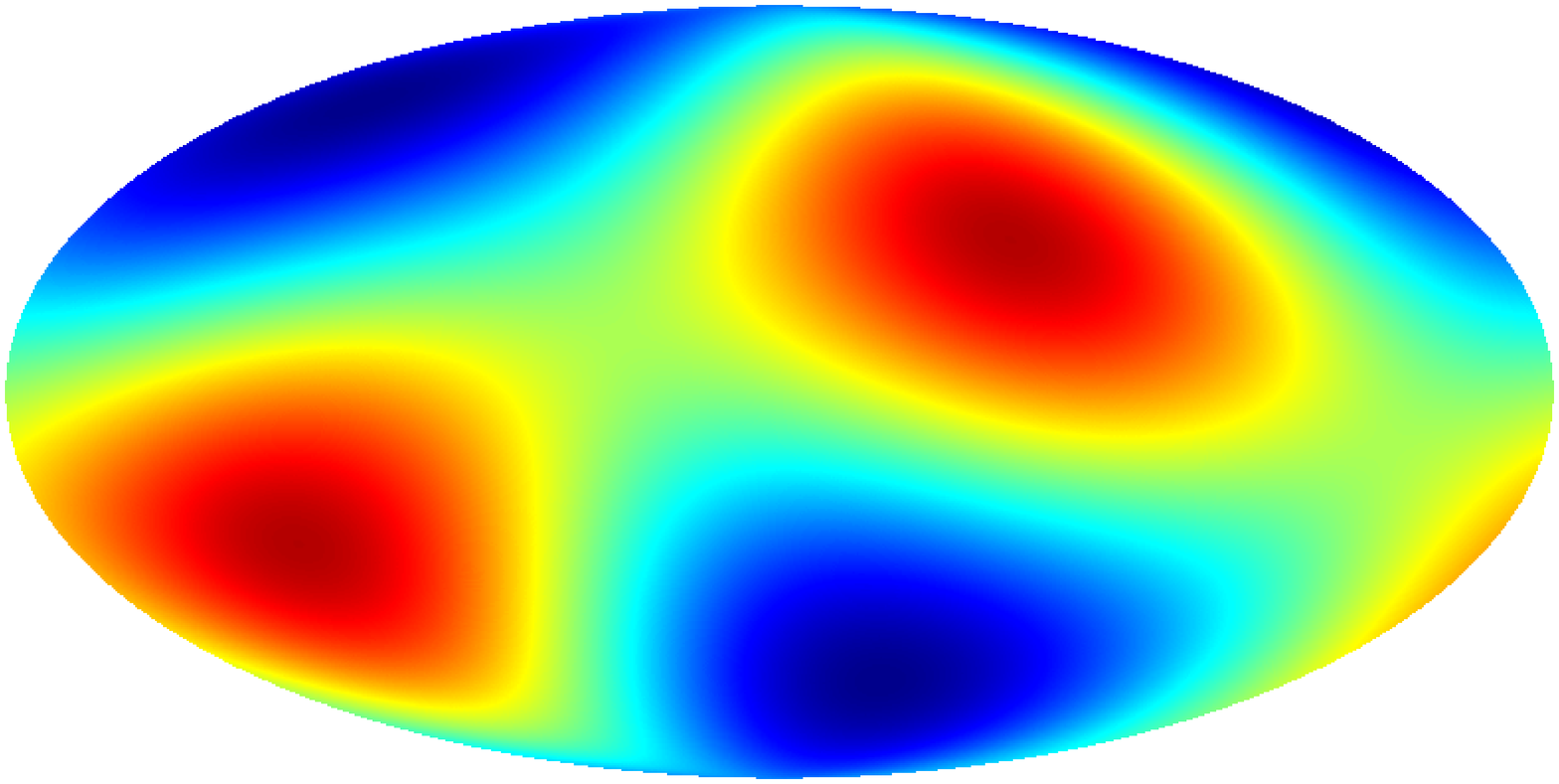}
\caption{{\it Left panel:} The quadrupole component of the Q-band foreground emission map estimated by the WMAP team, no mask is applied. {\it Right panel:} Same as the left panel, but the KQ75 mask has been applied. Both in Galactic coordinates.}
\label{fig:quad of foreground}
\end{figure}

\begin{figure}
\includegraphics[angle=90,width=0.22\textwidth]{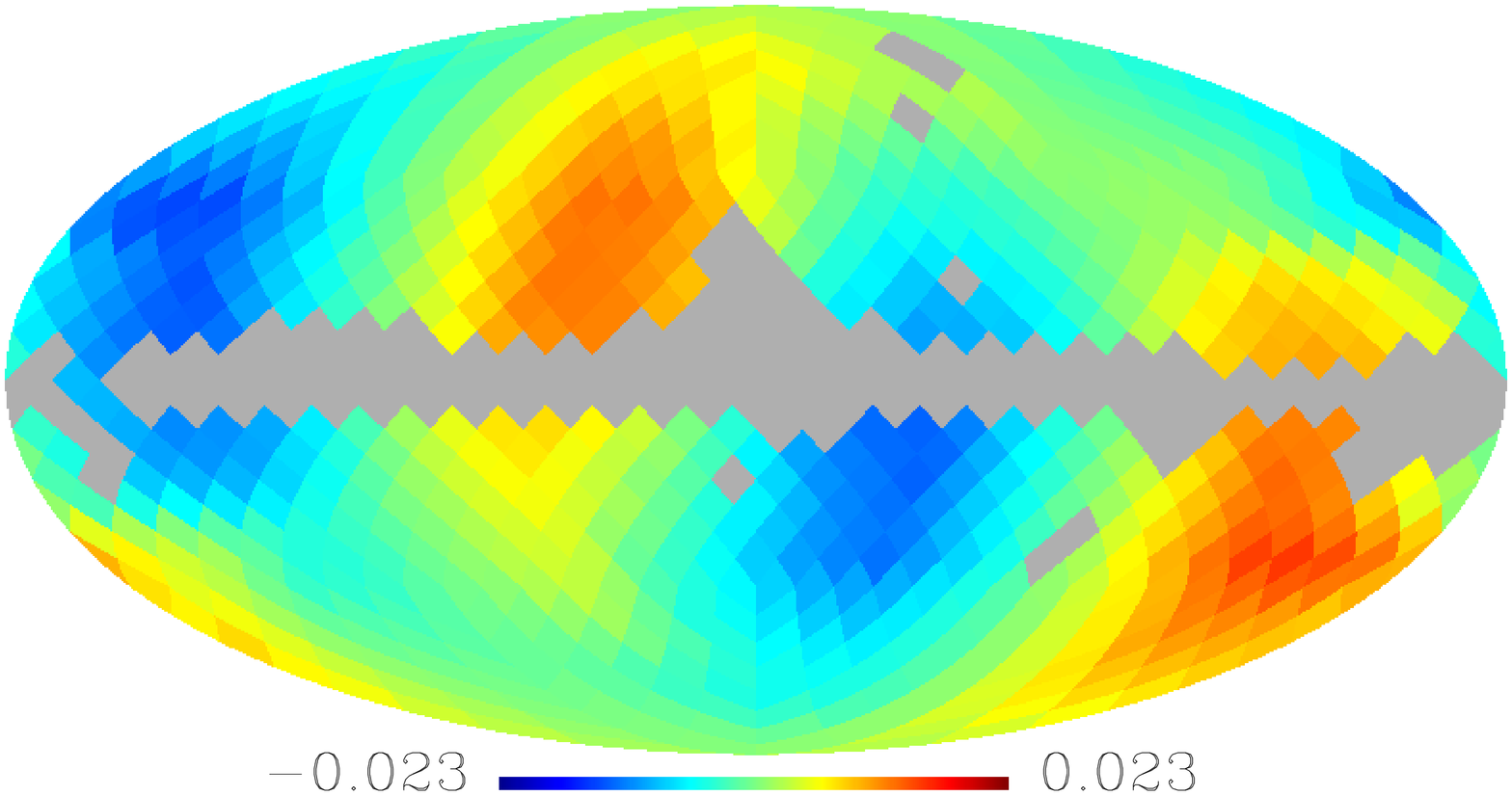}
\includegraphics[angle=90,width=0.22\textwidth]{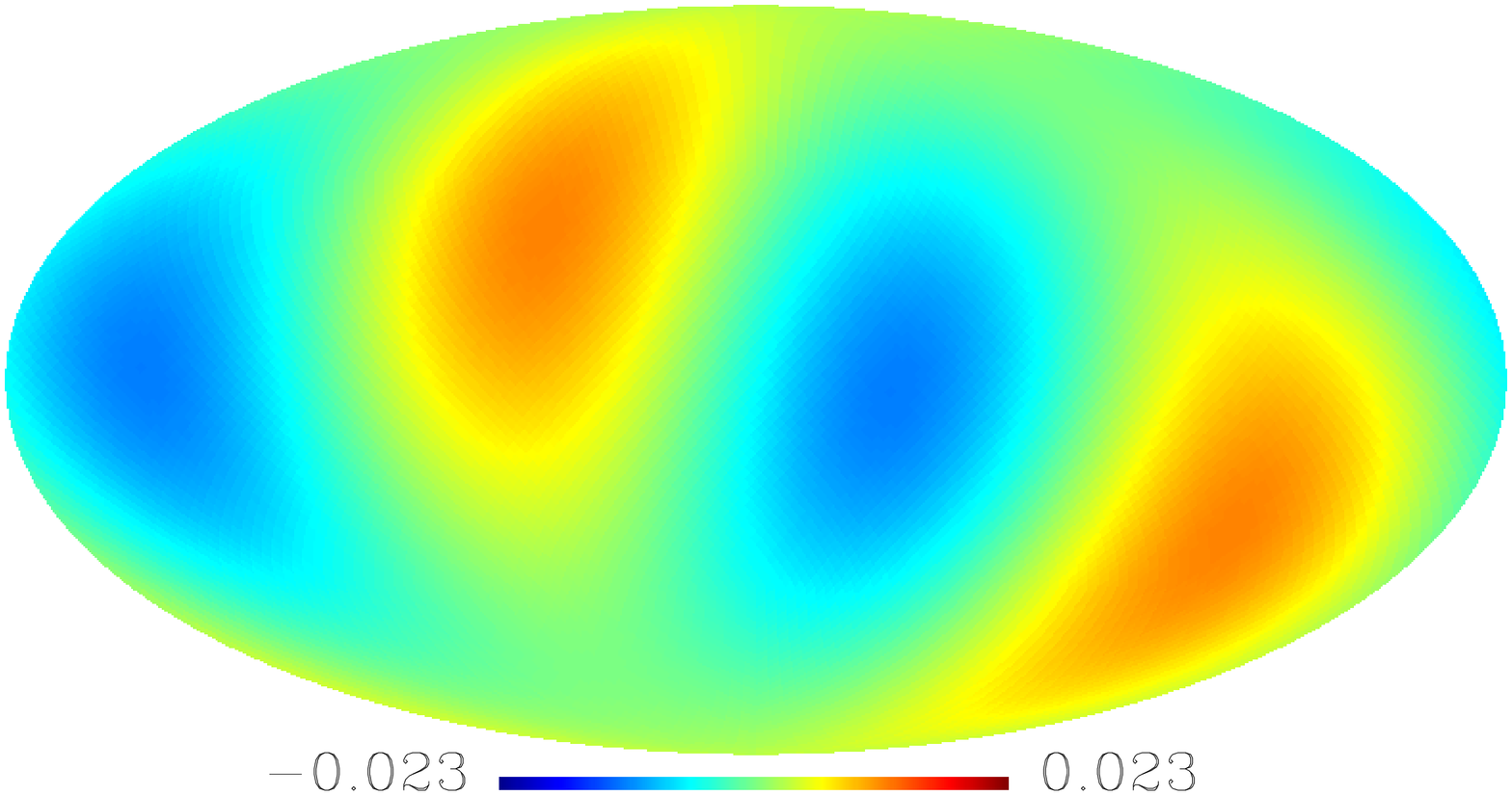}
\caption{{\it Left panel:} The artificial CMB quadrupole component produced by $-25.6$ ms timing offset. {\it Right panel:} The released WMAP CMB quadrupole component. Both in Galactic coordinate and units of mK.  Reproduced from Fig. 2 of our own work~\citep{lxl10}.}
\label{fig:compare}
\end{figure}

The diagnosed timing  error in this work might be due to coincidental correlation between the foreground emission and the quadrupoles in Fig.~\ref{fig:compare}  (no matter "real" or artificial).  Since the spherical harmonics $Y(l,m)$ with different $l$ or $m$ are exactly uncorrelated, this is same to the quadrupole of the foreground emission being correlated with Fig.~\ref{fig:compare}. However, as shown by Fig.~\ref{fig:quad of foreground}, the quadrupole component of the foreground emission is apparently uncorrelated with any one in Fig.~\ref{fig:compare}. Moreover, as shown in Table 1, the obtained $\<\Delta t^*\>$ is nearly frequency-independent. These two facts strongly suggests that we can safely ignore the foreground issue in this work.

It might also be worried that the true CMB quadrupole might be coincidentally correlated with the artificial quadrupole caused by the timing-asynchronism effect. In this case, the similarity we see in Fig.~\ref{fig:compare} is just something by chance. However, we have generated $10^8$ randomly distributed quadrupole pairs $(q_1, q_2)$ obeying the basic cosmic principles, especially the following two: There is no preferred axis, and there is no preferred spherical harmonic component (which means, the power expectations in $\rm{\mu K^2}$ for all $Y(l,m)$ components with $l=2$ and $ -2 \le m \le 2$ should be identical). For each pair, we calculate the correlation coefficient between $q_1$ and $q_2$, and with $10^8$ pairs we see that the probability of coincidentally getting a correlation coefficient greater than 0.8 (which is the correlation coefficient between the two panels in Fig.~\ref{fig:compare}) is about $2.5\%$. Thus we reject the assumption that the timing asynchronism effect we find in this work is due to coincidental correlation (in other words, the cross-term in Eq.~\ref{v}) at 0.025 confidence level.

It's also worthwhile to notice that, our work is done in the TOD space, which is significantly different to the temperature map space. Therefore, even if there were strong correlation in the temperature map space, it does not necessarily mean the same strong correlation in TOD space (the cross-term in Eq.~\ref{v}). This fact could further decrease the worry about the correlation issues.

\section{Conclusion}
\label{sec:conclusion}
In previous works we have found notable systematical errors in released WMAP temperature maps \citep{li09,liu09a}, which have been confirmed by other independent works (e.g. Aurich, Lustig \& Steiner 2009).  We then independently developed a self-consistent software package for WMAP data processing, and from the WMAP TOD produced new CMB maps  which are significantly different from the official maps (Liu \& Li 2009b, 2010a). Our pipeline codes are already publicly released on the website of Tsinghua Center for Astrophysics and the CosmoCoffee forum\footnote{Our software used for WMAP data processing has been thoroughly tested by us, and the result has been confirmed by other works as well (Moss, Scott \& Sigurdson 2010; Roukema 2010a). The software can be found at http://dpc.aire.org.cn/data/wmap/09072731/release\_v1/source\_code/v1/ or http://cosmocoffee.info/viewtopic.php?p=4525\#4525.}. Later, in searching for the source of the difference between our and official maps, we discovered a -25.6\,ms asynchronism between the spacecraft attitude and radiometer output timestamps in the official WMAP Meta and Science Data Tables, respectively; such a timing-offset, if not be properly corrected in data processing, should generate serious consequences in the recovered CMB map and power spectrum (Liu, Xiong \& Li 2010). We artificially introduced the -25.6\,ms asynchronism into our pipeline to simulate the WMAP manner and then obtained fully consistent result to the WMAP team indeed.

According to the well consistency between the -25.6\,ms timing-offset directly observed from TOD~\citep{lxl10} and indirectly probed by~\citet{rou10b} and this work, the most natural explanation should be existence of an unwanted timing error in the WMAP data. However, such an unwanted timing error is also expected to introduce a blurring effect in recovered maps, but~\citet{rou10a} did not find such effect in WMAP official maps. Recently, by comparing the median per map of the fluctuation variance per pix in the temperature map for different assumed timing-offsets,~\citet{rou10b} proved that there does exist an about -25.6\,ms timing-offset in the WMAP calibrated TOD, which is confirmed by us in this work directly in the WMAP TOD, and we further show that the uncorrected timing error occurred at least in calculating the Doppler dipole signal.  A natural explanation for above findings is that the timing-offset-induced error in direction did not have an effect in the compilation of the calibrated TOD into sky maps by the actual Jupiter pointing measurements \citep{hin03}, but the calibration error, which on large scales consists of a timing-offset-induced pseudo-dipole signal already present in the calibrated TOD, remains present in the sky maps.

It has to be noticed that the error in sky-map-based determinations of CMB dipole direction \citep{ben03b,hin09} can also contribute to the diagnosed timing-offset. In other words,  the timing-offset detected in this work is a synthesis of errors in timing and in dipole direction.  Besides the detected timing offset, other observational reasons, i.e. the sidelobe pickup contamination from dipole, can also generate artificial quadrupole aligned with what observed in the official CMB map, needed to be further removed by model fitting \citep{liu11a,liu11b}.  After template-based removal of artificial quadrupole, the remaining quadrupole power can be as low as $10.4\,\rm{\mu K^2}$, significantly lower than $28.6\,\mu$K$^2$ derived in this work, indicating that the sidelobe-pickup-induced artifact can not be ignored. The timing asynchronism in the WMAP raw data might be a problem special for the WMAP mission, whereas errors in dipole direction and sidelobe contamination are  common problems for all CMB missions.  To quantitatively estimate the effect of producing artificial CMB large-scale anisotropies from each possible source by thoroughly rechecking the WMAP data processing process is very important, not only for the WMAP-based cosmological study, but also for the Planck and other future CMB missions.

\acknowledgments We are greatly indebted to the anonymous referee for very detailed comments and helpful suggestions. This work is supported by the National Natural Science Foundation of China (Grant No. 11033003) and the National Basic Research Program of China (Grant No. 2009CB824800). The data analysis made use of the WMAP data archive (http://lambda.gsfc.nasa.gov/product/map/).

\end{document}